\documentclass[aps,prl,twocolumn,superscriptaddress,notitlepage,nofootinbib]{revtex4-1}

\usepackage{epsfig}
\usepackage{multirow}
\usepackage{slashed}
\usepackage{amsmath}
\usepackage{physics}
\usepackage{qcircuit}
\usepackage{braket}
\usepackage{graphicx}
\usepackage{subfigure}
\usepackage{color}
\usepackage{epstopdf}
\usepackage[inline]{enumitem}

\def\bea#1\eea{\begin{align}#1\end{align}}
\newcommand{\nn}{\nonumber\\}
\newcommand{\bef}{\begin{figure}[!htp]}
\newcommand{\eef}{\end{figure}}

\usepackage{xcolor}
\usepackage[normalem]{ulem}

\begin{document}

\title{Partonic collinear structure by quantum computing}

\date{\today}

\author{Tianyin Li}
\affiliation{Guangdong Provincial Key Laboratory of Nuclear Science, Institute of Quantum Matter, South China Normal University, Guangzhou 510006, China}
\affiliation{Guangdong-Hong Kong Joint Laboratory of Quantum Matter,
Southern Nuclear Science Computing Center, South China Normal University, Guangzhou 510006, China}

\author{Xingyu Guo}
\affiliation{Guangdong Provincial Key Laboratory of Nuclear Science, Institute of Quantum Matter, South China Normal University, Guangzhou 510006, China}
\affiliation{Guangdong-Hong Kong Joint Laboratory of Quantum Matter, Southern Nuclear Science Computing Center, South China Normal University, Guangzhou 510006, China}

\author{Wai Kin Lai}
\affiliation{Guangdong Provincial Key Laboratory of Nuclear Science, Institute of Quantum Matter, South China Normal University, Guangzhou 510006, China}
\affiliation{Guangdong-Hong Kong Joint Laboratory of Quantum Matter,
Southern Nuclear Science Computing Center, South China Normal University, Guangzhou 510006, China}
\affiliation{Department of Physics and Astronomy, University of California, Los Angeles, California 90095, USA}

\author{Xiaohui Liu}
\affiliation{Center of Advanced Quantum Studies, Department of Physics,
Beijing Normal University, Beijing 100875, China}
\affiliation{Center for High Energy Physics, Peking University, Beijing 100871, China}

\author{Enke Wang}
\email{wangek@scnu.edu.cn}
\affiliation{Guangdong Provincial Key Laboratory of Nuclear Science, Institute of Quantum Matter, South China Normal University, Guangzhou 510006, China}
\affiliation{Guangdong-Hong Kong Joint Laboratory of Quantum Matter,
Southern Nuclear Science Computing Center, South China Normal University, Guangzhou 510006, China}

\author{Hongxi Xing}
\email{hxing@m.scnu.edu.cn}
\affiliation{Guangdong Provincial Key Laboratory of Nuclear Science, Institute of Quantum Matter, South China Normal University, Guangzhou 510006, China}
\affiliation{Guangdong-Hong Kong Joint Laboratory of Quantum Matter,
Southern Nuclear Science Computing Center, South China Normal University, Guangzhou 510006, China}

\author{Dan-Bo Zhang} 
\email{dbzhang@m.scnu.edu.cn}
\affiliation{Guangdong-Hong Kong Joint Laboratory of Quantum Matter, Frontier Research Institute for Physics,
South China Normal University, Guangzhou 510006, China}  
\affiliation{Guangdong Provincial Key Laboratory of Quantum Engineering and Quantum Materials, School of Physics and Telecommunication Engineering,
South China Normal University, Guangzhou 510006, China}

\author{Shi-Liang Zhu} 
\email{slzhu@scnu.edu.cn}
\affiliation{Guangdong-Hong Kong Joint Laboratory of Quantum Matter, Frontier Research Institute for Physics,
South China Normal University, Guangzhou 510006, China}  
\affiliation{Guangdong Provincial Key Laboratory of Quantum Engineering and Quantum Materials, School of Physics and Telecommunication Engineering,
South China Normal University, Guangzhou 510006, China}

\collaboration{QuNu Collaboration}

\date{\today}         

\begin{abstract}
We present a systematic quantum algorithm, which integrates both the hadronic state preparation and the evaluation of real-time light-front correlators, to study parton distribution functions (PDFs). As a proof of concept, we demonstrate the first direct simulation of the PDFs in the 1+1 dimensional Nambu-Jona-Lasinio model. We show the results obtained by exact diagonalization and by quantum computation using classical hardware. The agreement between these two distinct methods and the qualitative consistency with QCD PDFs validate the proposed quantum algorithm. Our work suggests the encouraging prospects of calculating QCD PDFs on current and near-term quantum devices. The presented quantum algorithm is expected to have many applications in high energy particle and nuclear physics. 
\end{abstract}

\maketitle

\textit{Introduction}.---Identifying the fundamental partonic structure of hadrons has remained one of the main goals in both high energy particle and nuclear physics since the pioneering deep-inelastic scattering experiments at SLAC in the late 1960s~\cite{Lin:2017snn,Gao:2017yyd}. Such partonic structure is encoded in the universal parton distribution functions (PDFs), which represent the probability density to find a parton in a hadron with specified momentum fraction $x$ and resolution scale $\mu$. Besides characterizing the hadronic parton structure, accurate determination of the PDFs is mandatory for making predictions for hard scattering processes within the Standard Model and for providing the necessary benchmark information in the search for new physics beyond the Standard Model.

PDFs are nonperturbative quantities. They cannot be obtained from perturbative QCD calculations. Thanks to the QCD factorization theorems~\cite{Collins:1989gx}, PDFs can be extracted through global analysis of the experimental data. Extensive efforts have been devoted to this method; see, for instance, CTEQ~\cite{Hou:2019efy}, MMHT~\cite{Harland-Lang:2014zoa}, NNPDF~\cite{NNPDF:2017mvq}, AMBP~\cite{Alekhin:2017kpj}, HERAPDF~\cite{H1:2015ubc}, and JAM~\cite{Ethier:2017zbq,Barry:2018ort}. On the other hand, since PDFs are defined as real-time correlators of quark and gluon fields, their direct evaluations in Euclidean lattice QCD have encountered intrinsic difficulties for a long time. To circumvent this problem, there have been several proposals to obtain PDFs out of quantities that can be calculated directly by lattice simulations~\cite{Ji:2013dva, Ji:2014gla,Ma:2014jla, Ma:2017pxb,Radyushkin:2017cyf, Orginos:2017kos,Liang:2019frk,Chambers:2017dov,Sufian:2019bol,Izubuchi:2019lyk,Lin:2020ssv,Musch:2011er,Detmold:2001jb,LHPC:2007blg,Alexandrou:2015qia,Alexandrou:2020sml}.
% following the original idea of~\cite{Ji:2013dva}.
However, a direct calculation of the PDFs by evaluating the real-time light-cone correlators is still inaccessible by foreseeable classical computations.

Inspired by the great promise of quantum computers~\cite{Arute:2019zxq} and the remarkable success in various related fields~\cite{Alexeev:2019enj}, particularly in using the natural capability of quantum computation to simulate real-time evolution in quantum field theories, there has been a rapidly growing wave of interest in implementing quantum computing methodologies to particle and nuclear physics in recent years (see reviews~\cite{Cloet:2019wre, Zhang:2020uqo}).
The idea is to simulate hadron physics with well-controllable quantum devices by generating quantum states describing hadrons and evaluating physical properties through measurements on the hadronic states. A few pioneering attempts toward the calculation of PDFs on a quantum computer have been explored, such as the proposal of computing hadronic tensors~\cite{Lamm:2019uyc}, 
 quantum simulations of the space-time Wilson loops~\cite{Echevarria:2020wct}, 
the hybrid approach incorporating quantum computing as a subroutine~\cite{Mueller:2019qqj}, as well as global analyses of the world data with quantum machine learning~\cite{Perez-Salinas:2020nem}. Still, a comprehensive and unified framework for investigating PDFs on a quantum computer using the fundamental lattice theory approach is lacking, which calls for a systematic treatment, with concrete quantum algorithms as well as analyses of 
scalability. 

In this paper, we work out a systematic quantum computational approach for evaluating the PDFs. Specifically, we develop a framework of the variational quantum eigensolver for faithful preparation of hadronic states of given quantum numbers, with the advantage of efficient parametrization and almost trap-free optimization. As a conceptual proof of our algorithm, we realize, for the first time, a practical quantum simulation of the meson PDFs. For the purpose of illustration, we present the formulation in the $1+1$ dimensional Nambu-Jona-Lasinio (NJL) model~\cite{Nambu:1961tp,Nambu:1961fr}, also known as the Gross-Neveu (GN) model~\cite{Gross:1974jv}, which captures many fundamental features of QCD and provides a unified picture of the vacuum, mesons, and nucleons~\cite{Christov:1995vm}. We show in detail how to prepare the hadronic bound states efficiently on a quantum computer and how the real-time parton correlators in coordinate space can be evaluated directly. 
%The obtained meson PDF from quantum computing agrees qualitatively with those from conventional global fitting methods and lattice QCD calculations. 

\textit{Parton distribution functions and the NJL model}.---In order to illustrate the calculation of PDFs using quantum computing, we take the hadron PDFs in the $1+1$ dimensional NJL model as a concrete example to introduce our quantum algorithm. The Lagrangian for the NJL model is given by
\begin{equation}\label{LA}
	\mathcal{L}=\bar{\psi}_\alpha (i\gamma^\mu \partial_\mu-m_\alpha)\psi_\alpha +g(\bar{\psi}_\alpha \psi_\alpha)^2\,,
\end{equation}
where $g$ is the strong coupling constant and $m_\alpha$ is the quark mass, with $\alpha$ the flavor index. In the hadron rest frame, the quark PDFs can be written as
\bea\label{eq:pdf}
	f_{q/h}(x) =&\int \frac{dz}{4\pi} e^{-ixM_h  z}
	\nn 
&\times \bra{h} e^{iHt}\bar{\psi}(0,-z)e^{-iHt}\gamma^+ \psi(0,0)\ket{h} \,, \nn 
\eea 
where we have written the quark field ${\bar \psi}(zn^\mu)$ with the light-front coordinate $zn^\mu = z(1,-1)$ as ${\bar \psi}(zn^\mu) = e^{iHz}\bar{\psi}(0,-z) e^{-iHz}$ with $H$ the Hamiltonian of the NJL model, and we set $t = z$ in Eq.~(\ref{eq:pdf}).
Here $M_h$ is the mass of the hadron $h$. Throughout this work, the hadron state in the $h$ rest frame will be denoted by $\ket{h}$.

One of the major difficulties to directly compute the PDFs by classical lattice simulations is rooted in the evaluation of the real-time lightlike correlators in Eq.~({\ref{eq:pdf}}). In this work, we will manifest how the time-dependent correlators can be computed on a quantum computer in polynomial time.

For later evaluations of Eq.~(\ref{eq:pdf}) on a quantum computer, we discretize the space with $N/2$ lattice sites for each flavor $\alpha$ and put the quark fields on the lattice using the staggered fermion approach,
\bea
\psi_\alpha(x) =  
\begin{pmatrix}
\psi_{\alpha,1}(x) \\
\psi_{\alpha,2}(x) 
\end{pmatrix}	
%=
%\begin{pmatrix}
%\phi_{\alpha,\frac{(\alpha-1)N}{2}+2n} \\
%\phi_{\alpha,\frac{(\alpha-1)N}{2}+2n+1} 
%\end{pmatrix}
\equiv 
\begin{pmatrix}
\phi_{\alpha,2n} \\
\phi_{\alpha,2n+1} 
\end{pmatrix}
\,, 
\eea
with $0
\le n 
\le \frac{N}{2}-1$, 
%\old{where we distribute the up and down components of the Dirac spinor to different lattice} 
where we used $\phi$, instead of $\psi$, to denote the discretized up and down components of the Dirac spinor on the lattice, and we denoted and ordered the flavor indices $\alpha,\beta, \dots $ as $1,2,\dots$. 
We further apply the Jordan-Wigner transformation~\cite{backens_shnirman_makhlin_2019} to rewrite the fermionic fields in terms of the Pauli matrices that can be operated on quantum computers, 
\bea
\phi_{\alpha,n}  
= \prod_{\beta=1}^{\alpha-1}\tilde{\sigma}_{\beta,\frac{N}{2}}^3
\tilde{\sigma}_{\alpha,n}^3 \sigma_{\alpha,n}^+
\equiv 
\Xi_{\alpha,n}^3
\sigma_{\alpha,n}^+
\,, 
\eea
where we used
the raising and lowering operators $\sigma_{\alpha,n}^{\pm} = \frac{1}{2}( \sigma_{\alpha,n}^1 \pm i \sigma_{\alpha,n}^2)$ and introduced the string operator $\tilde{\sigma}^3_{\alpha,n}\equiv \prod_{i<n}\sigma^3_{\alpha,i}$ to simplify the notations. Here $\sigma^{j}_{\alpha,i}$ is the $j$th component of the Pauli matrix on the lattice site $i$ for flavor $\alpha$. Throughout the work, we always impose the periodic boundary condition on the lattice sites for each quark flavor $\alpha$. 
The PDFs are then found to be
\bea
		f_{q_\alpha/h}=\sum_z \frac{1}{4\pi}e^{-ixM_h z}D(z)\,, 
\eea
where the PDFs in position space are
\bea\label{eq:pdfstagger}
		D(z)=\sum_{i,j=0}^{1}
		\bra{h} e^{iHz}
        \phi^\dagger_{\alpha,-2z+i}
		e^{-iHz}
        \phi_{\alpha,j}
        \ket{h}\,.
\eea
Now we are ready to elaborate on the quantum algorithm we propose for the direct PDF evaluation.

\textit{Quantum Algorithms}.---The quantum algorithm we propose for calculating the PDFs consists of two parts:
\begin{enumerate*}
\item 
prepare the hadronic state 
using the quantum-number-resolving variational quantum eigensolver (VQE); 
\item calculate the dynamical light-front correlation function in Eq.~(\ref{eq:pdf}) or, equivalently, Eq.~(\ref{eq:pdfstagger}).
\end{enumerate*}

\begin{figure}[htbp]
	\centering
	\includegraphics[width=0.48 \textwidth]{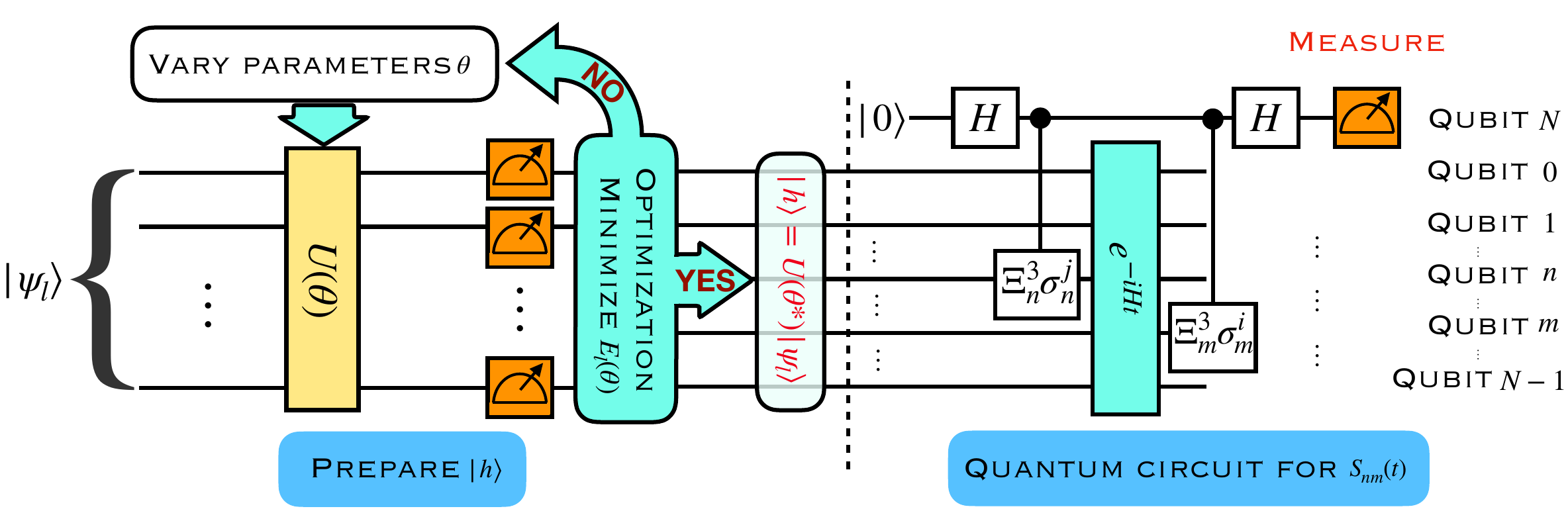}
	\caption{Quantum circuit for the calculation of PDFs. On the 
	left-hand side of the dashed line is the circuit for hadronic state preparation, while the right-hand side is for the correlation function. }
	\label{fig-circuit}
\end{figure}

{\it 1. Hadronic state preparation}.
Simulating hadronic states for a generic quantum field system on a quantum computer typically involves preparing excited states with specified quantum numbers. For this purpose, we develop a framework of the quantum-number-resolving VQE.
%\new{using the well established technique of VQE, which has been extensively used in a broad range of researches \cite{}}, 
%\old{with the advantage of efficient parametrization of hadronic states and almost trap-free optimization.} 
The quantum circuit for hadronic state preparation is illustrated on the left-hand side of the dashed line shown in Fig.~\ref{fig-circuit} and involves two steps.

First of all, we construct trial hadronic states for the $k$th excited state with a set of quantum numbers $l$ as
\begin{equation}
\ket{\psi_{lk}(\theta)}
= U(\theta) \ket{\psi_{lk}}_{\rm ref}\,,
\end{equation}
where $\ket{\psi_{lk}}_{\rm ref}$ is an input reference state and $U(\theta)$ encodes a set of symmetry-preserving unitary operators with parameters $\theta$. 

The input states $\ket{\psi_l}_{\rm ref}$ with discrete quantum numbers---such as the charge, the baryon number, and the spin---can be realized as computational bases specified by those quantum numbers. In addition, to preserve the spatial symmetries, the input state of a given momentum will be represented by a superposition of the computational bases, and its preparation is similar to those of the Dicke states~\cite{B_rtschi_2019}. Specifically for the PDFs, zero momentum is chosen for the trial state. 

The symmetry-preserving unitary evolution $U(\theta)$ is constructed using the quantum alternating operator ansatz~(QAOA~\cite{farhi2014quantum}, also related to the Hamiltonian variational ansatz~\cite{wiersema-PRXQuantum2020}). To retain the representation power yet preserve the quantum numbers for the ansatz, the Hamiltonian is divided as $H=H_1+H_2+\cdots + H_n$ with $n\ge2$, where every $H_i$ inherits the symmetries of $H$ and $[H_i,H_{i+1}]\not=0$. The $U(\theta)$ then consists of multiple layers, with each layer constructed by alternately evolving $H_j$ with parametrized time duration $\theta_{ij}$,
\setlength{\abovedisplayskip}{4.95pt}
\setlength{\belowdisplayskip}{4.95pt}
\bea
U(\theta) \equiv \prod_{i=1}^p \prod_{j=1}^n
\exp(i\,\theta_{ij}H_j)\,.
\eea
Here $p$ is the number of layers, and trial states with larger $p$ and $n$ represent the objective hadronic state more faithfully. 
By construction, the operator $U(\theta)$ will evolve the initial state $\ket{\psi_{lk}}_{\rm ref}$ into 
%\old{a different ray} 
$\ket{\psi_{lk}(\theta)}$ due to the noncommutativity of two consecutive evolutions $\exp(i\theta_{ij}H_j)$ and $\exp(i \theta_{ij-1}H_{j-1})$, while preserving the quantum numbers $l$. 

The second step is the optimization for hadronic states, which can be realized by
minimizing the cost function constructed out of a weighted combination of energy expectations~\cite{nakanishi_19},
\begin{equation}
	E_l(\theta)=\sum_{i=1}^k w_{li} \bra{\psi_{li} (\theta)}H\ket{\psi_{li} (\theta)} \,,
\end{equation}
for $k$ excited states with the same $l$
to determine the optimized parameter set $\theta^\ast$.
Here we require $w_{l1}>w_{l2}>\cdots>w_{lk}$. 
The $k$th excited hadron state $\ket{h}$ is then prepared as
$\ket{h} =
U(\theta^\ast)
\ket{ \psi_{lk} }_{\rm ref}$. 

Returning to the specific $1+1$D NJL model we are discussing, the mapping onto a quantum computer has been carried out in the previous section. The Hamiltonian is found to take the form 
\bea\label{spinham}
H = H_1 +  H_2 + H_3 + H_4\,,
\eea
where 
\bea
H_1 =& \sum_{\alpha}\sum_{n={\rm even}}^{\frac{N}{2}-1}
\frac{1}{4}\left(
\sigma_{\alpha,n}^1 \sigma_{\alpha,n+1}^2
- \sigma_{\alpha,n}^2 \sigma_{\alpha,n+1}^1
\right) \,, \nn
H_2 =&  \sum_\alpha \sum_{n={\rm even}}^{\frac{N}{2}-1} \, \frac{g}{2}\, 
\sigma_{\alpha,n}^3\sigma_{\alpha,n+1}^3
+ \sum_{\alpha,\beta} H_{\rm int,\alpha\beta} \,,  \nn
H_3 =& H_1(n={\rm even} \to n = {\rm odd}) \nn
& + \frac{1}{4} \sum_{\alpha} \tilde{\sigma}^3_{\alpha,\frac{N}{2}}\left(
\sigma_{\alpha,\frac{N}{2}-1}^2 \sigma_{\alpha,0}^1
- \sigma_{\alpha,\frac{N}{2}-1}^1 \sigma_{\alpha,0}^2 
\right)  \,, \nn
H_4 =& \sum_\alpha\sum_{n=0}^{\frac{N}{2}-1} \frac{m_\alpha}{2}(-1)^n(I - \sigma_{\alpha,n}^3)
- \frac{g}{2}(I - \sigma_{\alpha,n}^3)\,,
\eea
where the second line of $H_3$ comes from the periodic boundary condition, and the interaction between different quark flavors $\alpha$ and $\beta$ is given by
\bea
H_{{\rm int},\alpha\beta} & = 
- \frac{g}{2} \sum_{n={\rm even}}^{\frac{N}{2}-1}
\sigma_{\alpha,n}^3\sigma_{\beta,n}^3 
+ \sigma_{\alpha,n+1}^3\sigma_{\beta,n+1}^3  \nn 
& \quad \quad
-  \sigma_{\alpha,n+1}^3\sigma_{\beta,n}^3 
- \sigma_{\alpha,n}^3\sigma_{\beta,n+1}^3  \,.
\eea

Here, $H_2$ and $H_4$ are diagonal. It is also straightforward to check that $H_i$ satisfies the condition $[H_i,H_{i+1}] \ne 0$ and meanwhile inherits the symmetries of $H$, following from $[Q^i,H_i]= 0$ and $[T,H_i] = 0$ for $i= 1,\cdots, 4$. Note that $T$ is the even-lattice translation operator, acting as $T^{-1}\sigma_{n}^iT = \sigma_{n+2}^i$, and $Q^i$ include the flavors and the electromagnetic charge. It is noted that the mass $m_\alpha$ and the coupling constant $g$ can be absorbed into the parameter set $\theta$.

To prepare the input reference states for QAOA, we follow the strategy introduced before. For instance, to construct a zero-charge hadronic state with the lowest mass out of the $1$-flavor NJL model (the $1$st excitation state with respect to the vacuum $\Omega$), the input states can be chosen as
\bea
\ket{\psi_{\Omega,1}}_{\rm ref}
=& \ket{ 010101\dots 01} \,, 
\nn 
\ket{\psi_{\Omega,2}}_{\rm ref}
=& \frac{1}{\sqrt{N/2}}
\left(
\ket{1001,\dots,01} + \ket{0110,\dots,01} \right.\nn
& \left. 
+ \dots + \ket{0101,\dots,10}
\right)\,,
\eea
from which it immediately follows that $Q \ket{\psi_{\Omega,1}}_{\rm ref} = Q \ket{\psi_{\Omega,2}}_{\rm ref}$, where $Q = \frac{e_q}{2}\sum_n(I - \sigma_n^3)$ is the electromagnetic charge operator.

Then we apply the QAOA to evolve the reference states to the objective vacuum and the hadronic state by minimizing the cost function $E_\Omega(\theta)$. Within $E_\Omega(\theta)$, we can choose, for instance, $w_{\Omega,1} = 1$ and $w_{\Omega,2}= 0.5$. 
For the cases with more flavors, the hadronic states can be prepared in a similar way.

\textit{2. Dynamical correlation function}.
With the hadronic state at hand, to obtain the PDFs, we implement the quantum circuit illustrated on the right-hand side of the dashed line in Fig.~\ref{fig-circuit}, following the standard technique in Ref.~\cite{PhysRevLett.113.020505}, to evaluate the $2$-point dynamical correlation function, 
\bea\label{corr}
	S_{mn}(t)=\bra{h}e^{iHt}
	\Xi^3_m \sigma^i_m e^{-iHt}\Xi^3_n\sigma^j_n \ket{h}\,,
\eea
with $i(j) = 1,2$. 
The $D(z)$ in Eq.~(\ref{eq:pdfstagger}) can then be written as a sum of such correlation functions (see Supplemental Materials~\cite{supp}).

The quantum algorithm encodes the information of the dynamical correlation function into an ancillary qubit which controls and probes the system of the simulated hadron. More specifically, one inserts an evolution $e^{-i H t}$ between two $\Xi^3_n \sigma^j_n$ and $\Xi^3_m \sigma^i_m$ operations controlled by the ancilla and then measures the ancilla on the $\sigma^1$~($\sigma^2$) basis to get the real (imaginary) part of the correlation function. Assuming that the probability of getting $\ket{0}$ out of the ancilla is $p_{mn}(t)$ for a given $m$ and $n$ in the circuit, both the real and the imaginary parts of the correlation function can be found immediately through~\cite{PhysRevLett.113.020505}
\begin{equation}
  S_{mn}(t)=p_{mn}(t)-\frac{1}{2} \,,
\end{equation}
hence finishing the calculation of the PDFs. Here we also calculate the vacuum propagator by replacing $\ket{h}$ with the vacuum $\ket{\Omega}$ in Eq.~(\ref{corr}), and we follow~\cite{Collins:2011zzd,Jia:2018qee} to remove the disconnected contributions. 

We now estimate the time complexity of our quantum approach. The depth of the parametric quantum circuit for preparing the hadron state with the QAOA ansatz is expected to be $O(N)$~\cite{wiersema-PRXQuantum2020}. Since optimization with such ansatz is believed to be almost trapped-free and quick to converge~\cite{wiersema-PRXQuantum2020}, we could estimate the complexity of the state preparation as $O(N)$. Extracting the PDFs involves a series of measurements of the dynamical correlation function $S_{mn}(t)$ at $t=1,2,\cdots,N/2$, with each $t=n$ taking a cost of $O(n^2/\varepsilon)$ under a desired precision $\varepsilon$. The cost mostly comes from decomposing the real-time evolution $e^{iHt}$ with the Trotter formula~\cite{nielsen_chuang_2010}. From the above analysis, the total time complexity for calculating the PDFs is $O(N^3/\varepsilon)$.

The polynomial scaling with the number of qubits---and, more notably, the ability to attain real-time light-cone evolution---clearly shows a quantum advantage of calculating hadron PDFs on a quantum computer. Although justified only with a 1+1D quantum system, the quantum advantage could still hold for generic quantum systems in higher space-time dimensions and with gauge fields, if one can map the non-Abelian gauge field onto qubits in an efficient way~\cite{Brower:1997ha,Banerjee:2012xg,Stryker:2020sap}. Remarkably, even with the presence of a Wilson line in the correlator of Eq.~\eqref{eq:pdf} due to the gauge field, PDFs can still be efficiently evaluated on a quantum computer 
(see Supplemental Material~\cite{supp}).    

\textit{Results}.---Considering the current limitations of using real quantum devices, we present results using a classical simulation of the quantum circuit in Fig.~\ref{fig-circuit}. The simulation is performed on a desktop workstation with $16$ cores, using the open source packages QuSpin~\cite{quspin} and projectQ~\cite{Steiger2018projectqopensource}. For the time being, we use $N = 12$ or $18$ qubits and set $a = 1$ for the lattice spacing, which is sufficient to demonstrate the feasibility of the proposed quantum algorithm. In reality, the lattice spacing $a$ and quark mass $m$ will be fixed by the values of the coupling constant $g$ and relevant physical quantities such as hadron masses.

To verify the simulation, we first measure the mass of the lowest-lying $ud$-like hadron $M_{h}= \bra{h}H\ket{h}$ with respect to the vacuum energy $\bra{\Omega}H\ket{\Omega}$,  
in the NJL model with 2 flavors. We compare the result $M_{h,{\rm QC}}$ with $M_{h,{\rm NUM}}$, where the subscripts ${\rm QC}$ and ${\rm NUM}$ stand for the results obtained by quantum computing and exact diagonalization (ED), respectively. For simplicity, we set $m a = 0.2 $ for both flavors. The comparison is shown in Table~\ref{tabl-mass}. The differences between these two approaches are at the percent level, thus justifying the setups. It is also interesting to note that for small quark mass, the dominant contribution to the hadron mass comes from the interactions rather than the quark masses. This is consistent with the picture that QCD dynamics generates the majority of the hadron mass~\cite{Ji:1994av,Accardi:2012qut, Anderle:2021wcy,Yang:2018nqn,Ji:2021mtz}. 
\begin{table}[htbp]\label{twofcphmass}
	\begin{tabular}{|l|l|l|l|l|l|}
		\hline
		$g$               & 0.2      & 0.4      & 0.6      & 0.8      & 1.0      \\ \hline
		$M_{h,{\rm QC}} a$ & 1.002    & 1.810    & 2.674    & 3.534    & 4.352    \\ \hline
		$M_{h,{\rm NUM}} a$         & 1.001    & 1.801    & 2.659    & 3.509    & 4.342    \\ \hline
	\end{tabular}
	\caption{Calculated hadron mass as a function of $g$ with $N=12$. }
	\label{tabl-mass}
\end{table}
We also test the results with a heavier quark mass $m a = 0.8$ in the $1$-flavor model. In that case the hadron mass is dominated by the quark masses $2m$ from the hadron valence components. This is consistent with the expectation on the similarity with the mass configuration of a heavy quarkonium. 

Now we present the quark PDF of the lowest-lying zero-charge hadron in the $1$-flavor NJL model, which mimics the quark PDFs for $\pi^0$. Considering the reliability of the results and tolerable computing time, we use $N=18$ and fix the quark mass $m a = 0.8 $. We show distributions in both position space $D(z)$ and momentum space $f_q(x)$. 

\begin{figure}[htbp]
	\centering
	\includegraphics[width=0.45\textwidth]{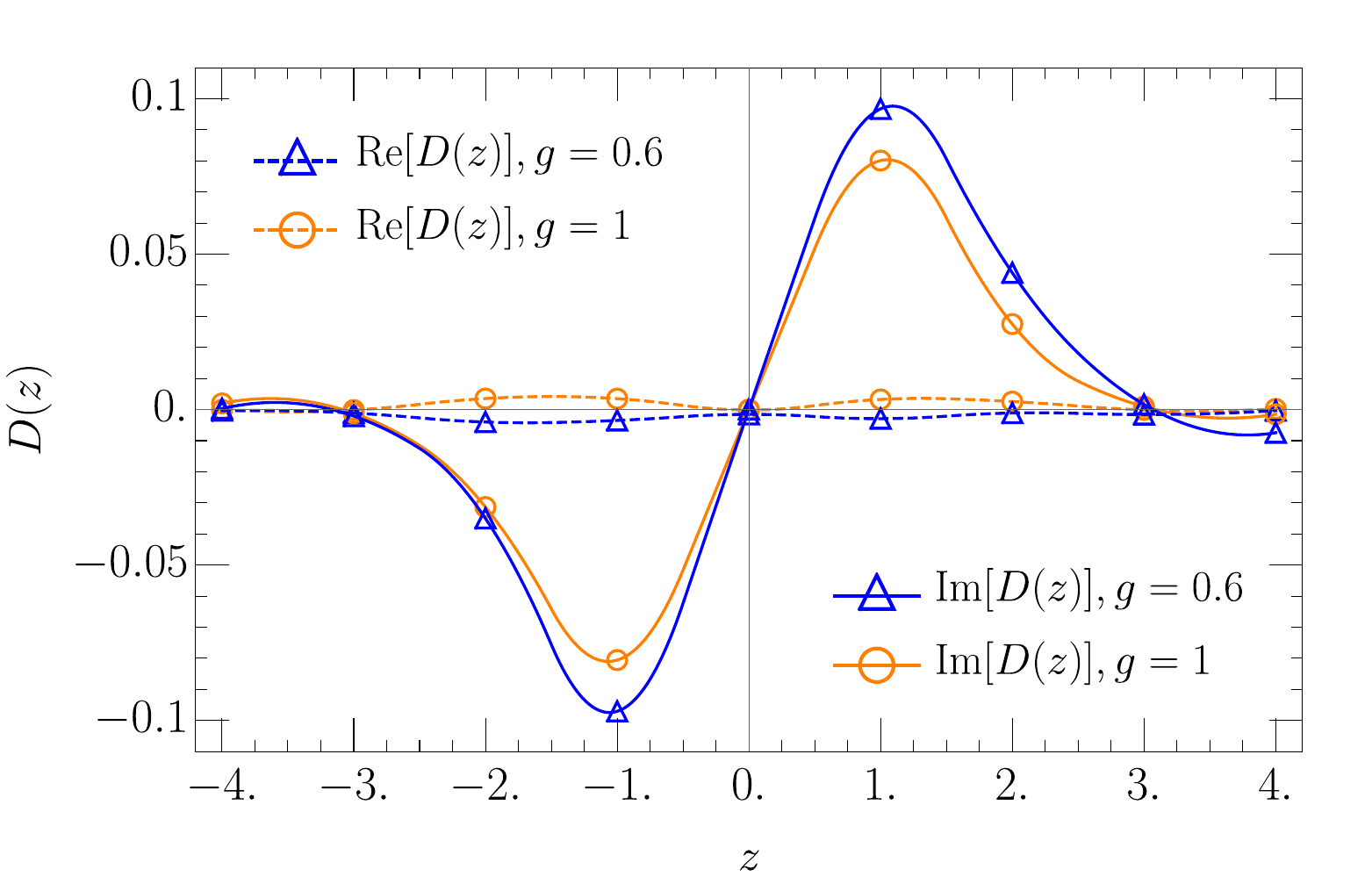}
\caption{Real (dashed lines) and imaginary parts (solid lines) of the quark field correlation function in position space. The discrete points are the lattice data, and the lines are from interpolations.
}
	\label{QCqusiN18sd}
\end{figure}
In Fig.~\ref{QCqusiN18sd}, we plot the imaginary and real parts of $D(z)$. We show results for both $g = 0.6$ and $g=1.0$. In the plot, 
the discrete open markers are from the direct quantum computations. %\textcolor{blue}{We notice that the calculated real part of $D(z)$ is consistent with zero.
%\footnote{\new{The small deviation from zero is mainly caused by finite volume effect.}} 
%With the fact that $f_q(x)$ is real, this implies that $f_q(x) = - f_q(-x)$. Since the PDF we calculate has the disconnected part subtracted away, the identity $f_{\bar q}(x) = -f_q(-x)$ holds~\cite{Collins:2011zzd} and thus we have $f_{q}(x) = f_{\bar q}(x)$, as expected for a $q\bar{q}$ bound state.} 
The real part shown in Fig.~\ref{QCqusiN18sd} is compatible with zero, implying that $f_q(x)=-f_q(-x)$.
Since $f_{\bar{q}}(x)=-f_q(-x)$ holds after subtraction of the disconnected part~\cite{Collins:2011zzd}, we have
$f_q(x)=f_{\bar{q}}(x)$, which is consistent with our expectation for one-flavor PDFs. 
To obtain the PDF $f_q(x)$, a continuous Fourier transformation of $D(z)$ will be performed. For this purpose, we interpolate the discrete results, as shown by the lines in Fig.~\ref{QCqusiN18sd}.% \textcolor{red}{where we observe similar shapes for both real and imaginary parts as those from Euclidean lattice calculations for pion PDFs \cite{Izubuchi:2019lyk} and proton PDFs \cite{Chen:2017mzz}.}

\begin{figure}[htbp]
	\centering
	\includegraphics[width=0.45\textwidth]{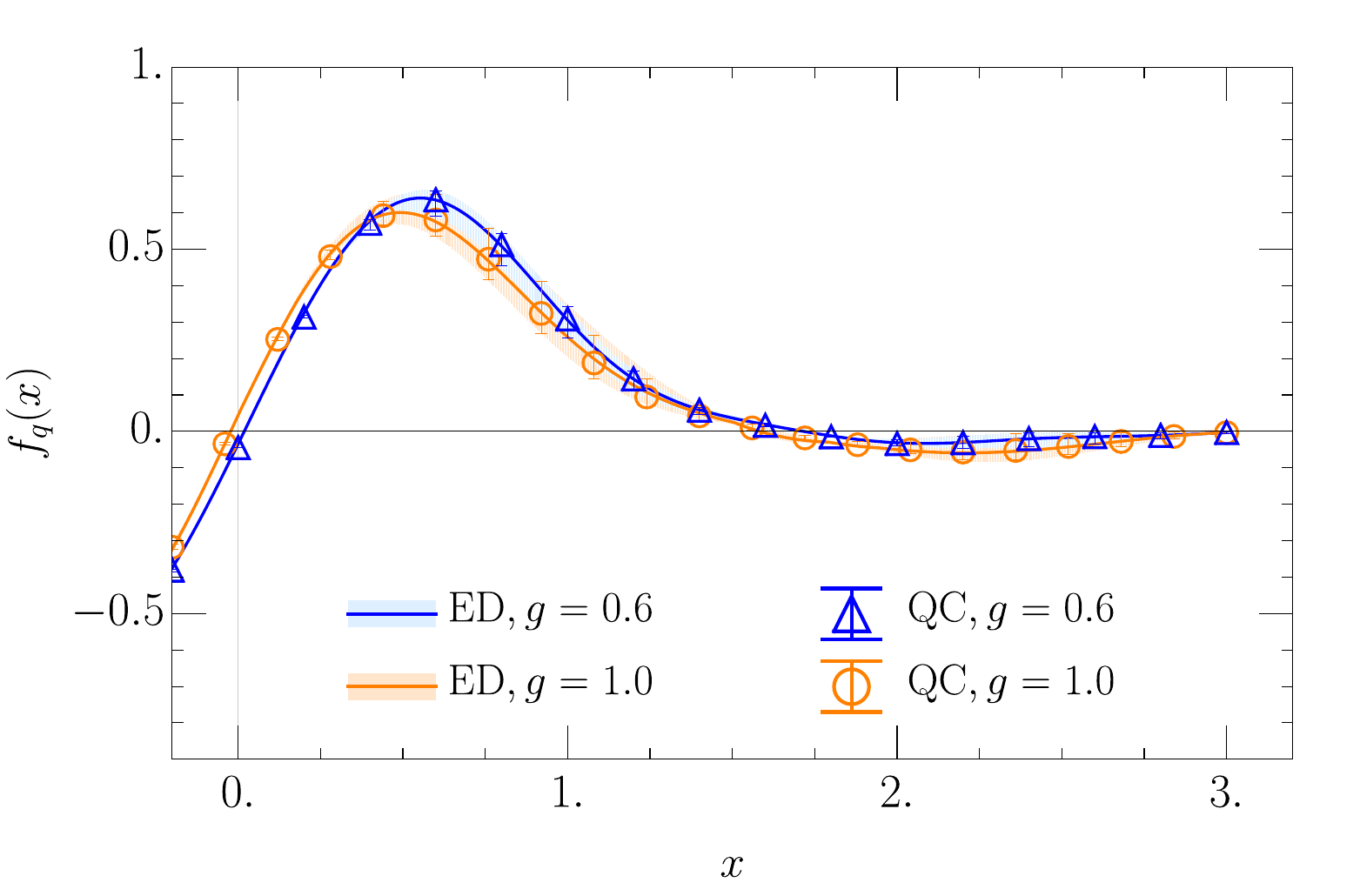}
	\caption{Quark PDF from quantum computing (open markers) and ED (solid lines). The error bars/bands arise from the estimated uncertainties due to different interpolation methods.
	}
	\label{QCPDFsN18}
\end{figure}

The results for the PDF are given in Fig.~\ref{QCPDFsN18}. We show $f_q(x)$ from both the  quantum computations (in open markers) and the ED to the NJL model (in solid lines). A perfect agreement between these two distinct approaches is observed. In Fig.~\ref{QCPDFsN18}, the error bars/bands illustrate the estimated uncertainties due to different ways of interpolations in $D(z)$. The nonvanishing but suppressed contributions in the nonphysical region $x > 1$ are partly due to the finite volume effect, which is also commonly seen in lattice calculations (see, for instance, Ref.~\cite{Ishikawa:2019flg}). It is expected to be further suppressed when more lattice sites and hence a larger volume are used. 
%In the small $x$ regime, we observed a growth in the PDF as $g$ increases, which indicates the increasing probability of creating more $q ({\bar q})$ out of the vacuum and thus lowering the momentum fraction when the interaction becomes strong. 
Here we note that the renormalization and calibration of input parameters, such as the lattice spacing $a$ and the quark mass $m$, for the NJL model are not yet considered in this initial exploration, whose focus is the practicability of a generic quantum algorithm for PDF evaluations. A discussion of the renormalization procedure and specific tuning of input parameters to physical quantities for the NJL model and QCD will be left for future publications. %\sout{Nevertheless, from Fig.~\ref{QCPDFsN18}, we observe qualitative agreements between the shapes of the calculated $ f_q(x)$ in $1+1$D NJL model and the real pion valence distribution from the global fitting~\cite{Barry:2018ort} \textcolor{red}{and the Euclidean lattice calculations~\cite{Sufian:2019bol,Izubuchi:2019lyk,Lin:2020ssv}. }} 
Nevertheless, 
%\old{from Fig.~\ref{QCPDFsN18}, we observe the expected peak around $x=0.5$ and qualitative agreements between the calculated $f_q(x)$ and the pion PDFs in two-dimensional QCD~\cite{Jia:2018qee}} 
we observe the expected peak around $x=0.5$, and the shapes shown in Fig.~\ref{QCPDFsN18} are in qualitative agreement with the calculated $f_q(x)$ in two-dimensional QCD~\cite{Jia:2018qee} and the extracted pion PDFs from the JAM Collaboration \cite{Barry:2021osv}.

\textit{Summary}.---In this work, we presented the first direct simulation of the hadron partonic structure on a quantum computer.
To realize such a calculation, we proposed a systematic quantum algorithm for preparing hadronic states using the quantum-number-resolving VQE, and we designed the quantum circuit to perform the real-time evolution of the lightlike correlators. As a proof of concept, we demonstrated the viability of the quantum algorithm by calculating the PDF in the $1+1$ dimensional NJL model. Our simulation results using only $18$ qubits agree with the ED to the discretized NJL model. 
%\sout{and remarkably, the simulated hadron PDFs \textcolor{red}{exhibit features}  qualitatively consistent with those from global fitting}. 
Our results manifest the quantum advantage over the classical Euclidean lattice methodology in resolving the intrinsic difficulties of simulating real-time dynamics. Our exploratory study suggests the possible future capability of studying realistic hadron structures on current and/or near-term quantum computers. %\old{with more than $50$ qubits}. 
Lastly, we point out that the proposed quantum computing framework for preparing the hadronic states and measuring the dynamical correlation functions is generally applicable. Its extension to many applications in high energy particle and nuclear physics is expected, such as calculations of scattering processes and parton showers~\cite{Bepari:2020xqi,Bauer:2019qxa,Bauer:2021gup}, the proton spin configurations, the 3D structure of a nucleon, as well as the fundamental jet transport properties~\cite{Barata:2021yri} and the associated dynamics~\cite{Cohen:2021imf,deJong:2020tvx} of quark-gluon plasma.

~~~~~
\begin{acknowledgments}
\textit{Acknowledgements}.---We thank Xiangdong Ji, Yu Jia, Jian Liang, Yan-Qing Ma, Wei Wang, Xiaonu Xiong and Jian-hui Zhang for helpful discussions. This work is supported by the Guangdong Major Project of Basic and Applied Basic Research No. 2020B0301030008, the Key-Area Research and Development Program of GuangDong Province (Grant No. 2019B030330001), the Key Project of Science and Technology of Guangzhou (Grant No. 2019050001), the National Natural Science Foundation of China under Grants No. 12022512, No.~12035007 (H.X.), No.~11775023 (X.L.), No.~12005065 (D.B.), and No.~12074180(S.L.), and by the Guangdong Basic and Applied Basic Research Fund No.~2021A1515010317 (D.B.). 
\end{acknowledgments}

%%%%%%%%%%%%%%%%%%%%%%%%%%%%%%%%%%%%%%%%%%%%%%%%%%%%%%%%%%%%

%\bibliographystyle{h-physrev5}   
%\bibliography{refs.bib}

\end{document}